# Parametric Analysis of First High-Gain Vertical Fe-doped Ultrafast Ga$_2$O$_3$ Photoconductive Semiconductor Switch

N. Karpourazar, *Student Member, IEEE*, S. K. Mazumder, *Fellow, IEEE,* V. Jangir, *Student Member, IEEE,* K. M. Dowling, J. Leach, and L. Voss

*Abstract*—**We investigate, as part of a Lawrence-Livermore-National-Laboratory (LLNL) sponsored-research work initiated in February of 2021, the parametric performance analysis of ultra-wide bandgap (UWBG) Fe-doped Ga$_2$O$_3$ photoconductive semiconductor switch (FG-PCSS) with embedded electrode. The detailed SILVACO based simulation of the FG-PCSS uses experimentally obtained lifetime, absorption coefficient, and mobility data. The key analysis results, demonstrated first to the sponsor LLNL in 2022, focus on the performance of the FG-PCSS under relatively high electric field and optical-excitation energy with specific regard to current gain, quantum efficiency, on-state resistance, and impact of beam position. The parametric analysis indicates that a high-gain operation yielding a low-cost laser beam for the FG-PCSS is possible.**

*Index Terms*— **Fe-doped Ga$_2$O$_3$, photoconductive, switch, avalanche, optical, gain, electric field**

## I. Introduction

UWBG material Ga$_2$O$_3$ (E$_G$ ~ 4.8 eV) has a high critical electric field (E$_c$ ~ 8 MV/cm), which yields devices with certain improved performance compared to SiC/GaN [1][2]. It can be grown from a melt and produced with a high crystalline quality at low cost as compared to (bulk) GaN, SiC, AlN, and diamond [3]. However, optical triggering of intrinsic UWBG (and WBG) PCSS requires high-power short-wavelength laser that is expensive and not readily available. For WBG PCSS, sub-bandgap triggering, using extrinsic semiconductors, has been proposed as a viable approach. For instance, in [4], Fe-doped GaN PCSS has been triggered using ~500 nm wavelength laser and operational feasibility demonstrated under linear mode of optical activation. However, the quantum efficiency of the extrinsic WBG device is relatively low since the absorption depth of the device due to mid wavelength optical excitation is on the higher side. Thus, for higher current operation, expensive high-power laser is required.

To address this apparent deficiency, researchers have explored the operation of PCSS in the high-gain nonlinear mode [5]. However, for narrow-bandgap and WBG PCSS, for the same power bias, the dimension of the PCSS is higher than that expected using UWBG PCSS such as FG-PCSS due to the higher breakdown electric field of $Ga_2O_3$. As such, this letter focuses on the performance of the FG-PCSS under relatively high electric field and low optical-excitation energy with specific regard to current gain, quantum efficiency, on-state resistance, and impact of beam position.

## II. FG-PCSS Device Outline

*A) Device Structure*

The FG-PCSS structure is shown in Fig. 1 with the anode (cathode) connected to bias (ground). The $Ga_2O_3$ material has a nominal defect (donor trap) concentration of $1\times10^{16}$ cm$^{-3}$ located at 0.56 eV from the conduction band, which is compensated by Fe with an acceptor trap concentration of $1\times10^{17}$ cm$^{-3}$ located at 0.76 eV from the conduction band. The device structure consists of two embedded ohmic electrodes of 1 micron thickness with the contact separation parametrically varied between 10 and 30 microns. For sub-bandgap optical triggering, a 450 nm wavelength is used, and the illumination is assumed to be uniform with the same optical intensity. A 400-micron device thickness ensures reasonable capture of photons.

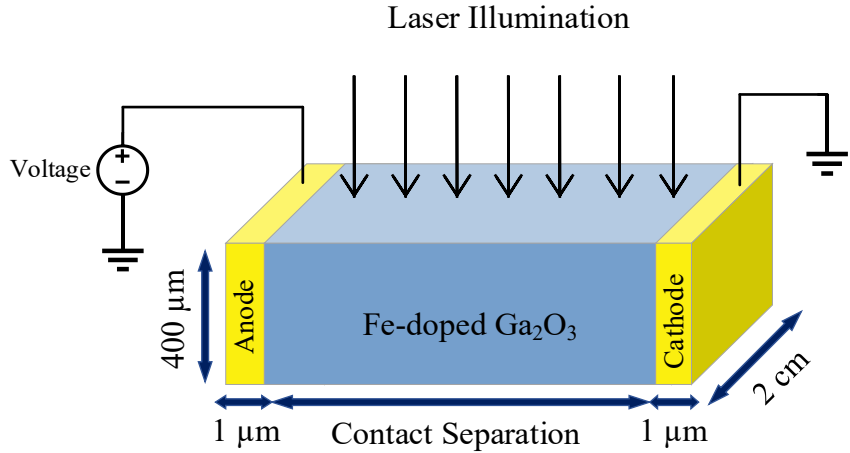

Fig. 1. Device structure of the FG-PCSS

*B) Empirical Determination of Device Material Parameters*

Beyond the device geometry, experiments were conducted to determine the absorption coefficient and the carrier lifetime.

A deuterium lamp and a StellarNet EPP200 spectrometer were used to experimentally determine the absorption spectrum of a Fe-Ga$_2$O$_3$ sample. Subsequently, using the experimentally obtained transmittance results, absorption coefficient, determined using (1) [6]:

$$\alpha = (-\ln T)/d \tag{1}$$

was found to be 2.2 cm$^{-1}$ for a wavelength of 450 nm. In (1), $\alpha$ is the absorption coefficient, $T$ is transmittance, and $d$ is the thickness of the Fe-Ga$_2$O$_3$ sample.

Lifetime was experimentally characterized in Fe-Ga$_2$O$_3$ using a setup described in [7]. A 30 ps laser pulse (355 nm) was applied on a coplanar waveguide structure with a gap in the signal trace. A large dc bias was applied and the conductivity across the gap was measured by detecting the current across the 50 Ω load in a 33 GHz oscilloscope (Keysight MSAV334A). The decay in the detected waveform corresponds to the recombination time of free electrons back into the deep acceptor states (Fe$^{3+}$/Fe$^{2+}$) [8], which was extracted from the slope of a logarithmic fit of the decay tail. From these measurements we found the carrier lifetime to be around 100 ± 5 ps for this material. Our simulations proceeded using 100 ps.

For low-to-moderate electric field, an electron mobility of 107 cm$^2$/(V.s) is used based on the Hall measurement. The hole mobility is set to be 20 cm$^2$/(V.s) based on [9]. Further, for higher electric field, the choice of electron and hole mobilities are guided by field-dependent mobility model of SILVACO as captured in [10] and was compared against the DFT-based results reported in [11]. The thermal conductivity of Ga$_2$O$_3$ is ~0.2 W/(cm.K) [3].

### III. Mathematical Model for FG-PCSS

The mathematical modeling of the FG-PCSS is based on the ; numerical solution of the semiconductor transport equations, which include the effect of the drift and diffusion currents along with the generation and the recombination terms following (2):

$$\frac{\partial n}{\partial t} = \frac{1}{q} div \vec{J}_n + G - R, \quad \frac{\partial p}{\partial t} = -\frac{1}{q} div \vec{J}_p + G - R. \tag{2}$$

In (2), $n$ $(p)$, $\vec{J}_n$ $(\vec{J}_p)$, $G$, $R$, and $q$ represent, respectively, the electron (hole) concentrationn, current density, generation rate, recombination rate, and the value of electron charge.

For $R$ in (2), we consider SRH, radiative, and Auger recombinations. Due to the presence of the trap levels, the standard SRH recombination rate is modified as follows [12]:

$$R_{SRH} = \sum_{i=1}^{l} R_{n_i} + \sum_{j=1}^{m} R_{p_j} \tag{3}$$

where $l$ $(m)$ represents number of donor (acceptor) traps and

$$R_{n,p} = \frac{pn - n_i^2}{\tau_n \left[ p + \frac{1}{g_0} n_i \exp \frac{E_i - E_T}{kT_L} \right] + \tau_p \left[ n + g_0 n_i \exp \frac{E_T - E_i}{kT_L} \right]}. \tag{4}$$

In (4), $\tau_n$ and $\tau_p$ are electron and hole lifetimes, $g_0$ is the degeneracy factor of the trap, $E_T$ is energy level of a trap, and $n_i$ is intrinsic carrier concentration, which is calculated by SILVACO using electron and hole effective mass captured from [13]. The radiative-recombination rate is captured by (5):

$$R_{rad} = C_{rad}(np - n_i^2) \tag{5}$$

where the $C_{rad}$ is the radiative-recombination coefficient [14]. Finally, Auger-recombination rate is obtained using

$$R_{Auger} = \alpha_n(pn^2 - nn_i^2) + \alpha_p(np^2 - pn_i^2) \tag{6}$$

where $\alpha_n$ and $\alpha_p$ are the Auger-recombination coefficients.

For $G$ in (2), Selberherr's ionization-rate model is used that is a variation of the classical Chynoweth model [15]:

$$G = \beta_n |\vec{J_n}| + \beta_p |\vec{J_p}| \tag{7}$$

where $\beta_n$ ($\beta_p$) (see (8)) and $\vec{J_n}$ ($\vec{J_p}$) represent electron (hole) ionization coefficient and current density:

$$\beta_n = A_n e^{(-B_n/E)^{\gamma_n}}, \beta_p = A_p e^{(-B_p/E)^{\gamma_p}} \tag{8}$$

In (8), $A_n$, $A_p$, $B_n$, $B_p$, $\gamma_n$, and $\gamma_p$ are user defined [16] and $E$ is the electric field in the direction of current flow. The ATLAS tool of SILVACO was used for simulating FG-PCSS. The device configured generated ~31,920 mesh points, requiring on an average, a simulation time of about 15 hours.

## IV. SIMULATION RESULTS

Fig. 2(a) shows the photocurrent response of the FG-PCSS (to a 10 μJ, 450-nm laser) as a function of the applied voltage bias for 10 μm and 30 μm contact separations (i.e., optical window lengths). For the 10 μm contact separation, the photocurrent of the FG-PCSS increases gradually up to 2 kV bias and increases exponentially beyond that operating point. The exponential increase in the FG-PCSS photocurrent onsets due to impact ionization at high electric field that yields avalanche multiplication. This is corroborated by observing that for the FG-PCSS with a 30 μm contact separation, similar nonlinear response in the photocurrent is not observed for the range of the voltage bias. Fig. 2(b) shows the FG-PCSS photocurrent response with 10 and 30 μm contact separations as a function of the laser energy up to 100 µJ at a fixed wavelength of 450 nm, for a linear and a nonlinear region, following Fig. 2(a). Fig. 2(b) demonstrates that increase in optical energy does not change the mode of operation of the FG-PCSS operating in either of the two regions, thereby indicating that the transition between the linear and nonlinear regions of operation in Fig. 2(a) is due to increasing electric field (owing to increasing voltage bias) that yields higher impact ionization.

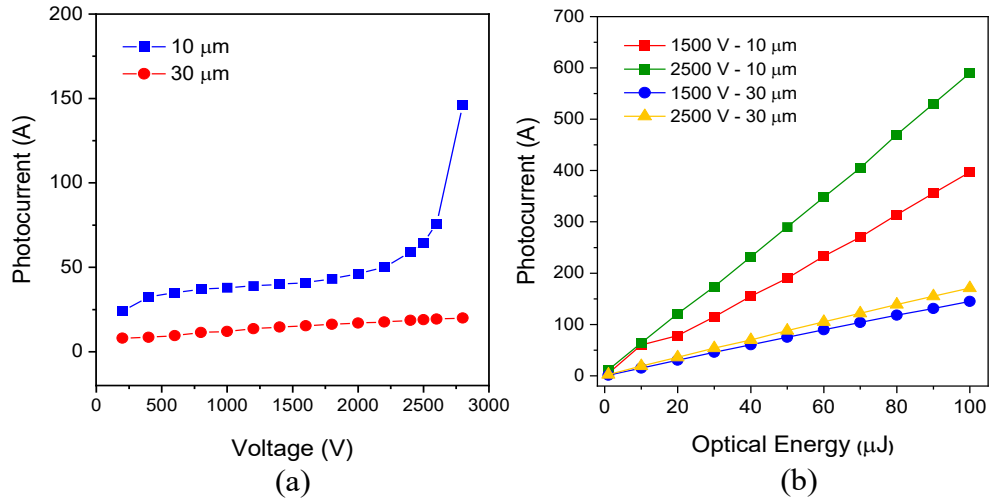

Fig. 2. FG-PCSS photocurrent verses (a) voltage, (b) optical energy.

Fig. 3 shows the electric field and impact ionization for a 10 µm contact separation at 200 V and 2500 V bias. The electric field is found to be high near the cathode and with increasing voltage, this high electric field extends to the anode due to charge accumulation. Impact ionization at 2500 V is found to be ~10 orders higher than that at 200 V.

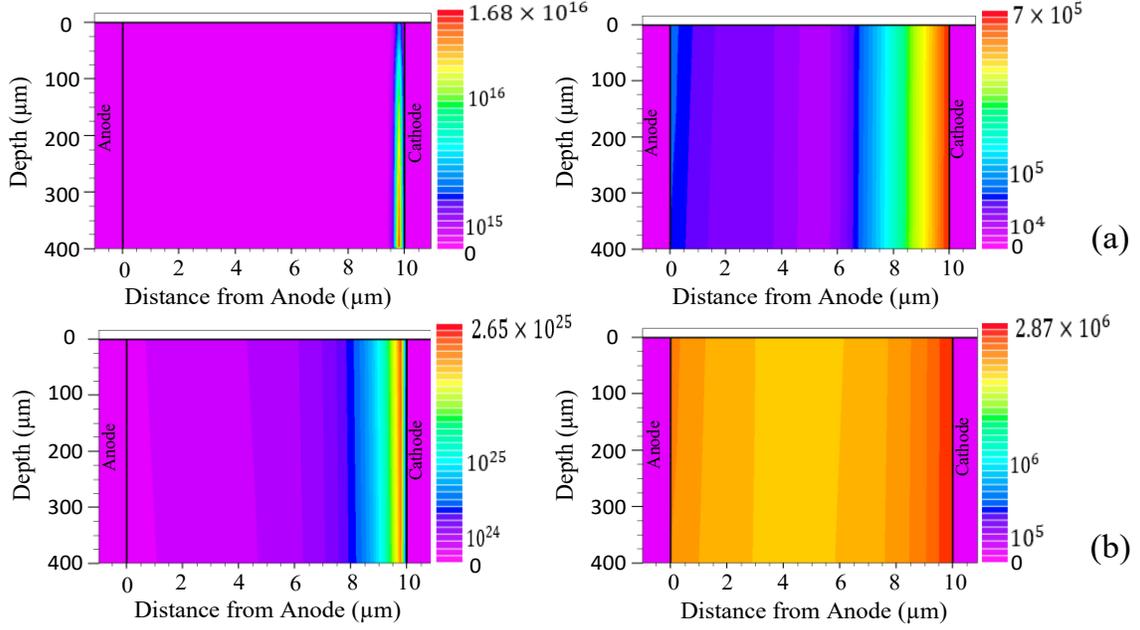

Fig. 3. FG-PCSS (left) impact ionization (cm$^{-3}$s$^{-1}$) and (right) electric field (V/cm) within the contact separation at a bias of (a) 200 V, (b) 2500 V.

Fig. 4 shows the on-resistance of the FG-PCSS with 10 µm and 30 µm contact separation as a function of the laser energy, using a 450-nm laser for biases of 1500 V and 2500 V. There is no significant difference between the current under 1500 V and 2500 V bias for a contact separation of 30 µm, and the on-resistance at 2500 V is slightly higher than that at 1500 V in the same situation. When the recombination time of the FG-PCSS material is much lower than the optical-pulse duration, the photocurrent follows the light pulse.

In this case, the PCSS on-resistance is given by [17]:

$$R_c = h_s^2 E_o^{-1}\left(E_p/(1-r)\,q\mu_s T_r\right) \qquad (9)$$

In (9), $h_s$ is the contact separation of the FG-PCSS, $E_p$ is the energy of a photon, $r$ is the surface-reflection coefficient, $\mu_s$ is the sum of the hole and electron mobilities, $T_r$ is the recombination time, and $E_o$ is the total optical energy. Equation (9) shows that the FG-PCSS with a smaller contact separation yields a lower on-resistance, necessitating reduced optical energy. The latter also enables the usage of a laser with shorter rise and fall times, which enhances the switching speed.

Fig. 5(a) shows the rise time with respect to incident beam energy at fixed applied voltage of 2500 V and Fig. 5(b) indicates rise time variation with applied voltage at fixed optical beam energy (10 μJ). The FG-PCSS has sub nano-second rise time which allows the device to operate at higher switching frequency. The rise time improves with higher optical beam energy and applied bias voltage.

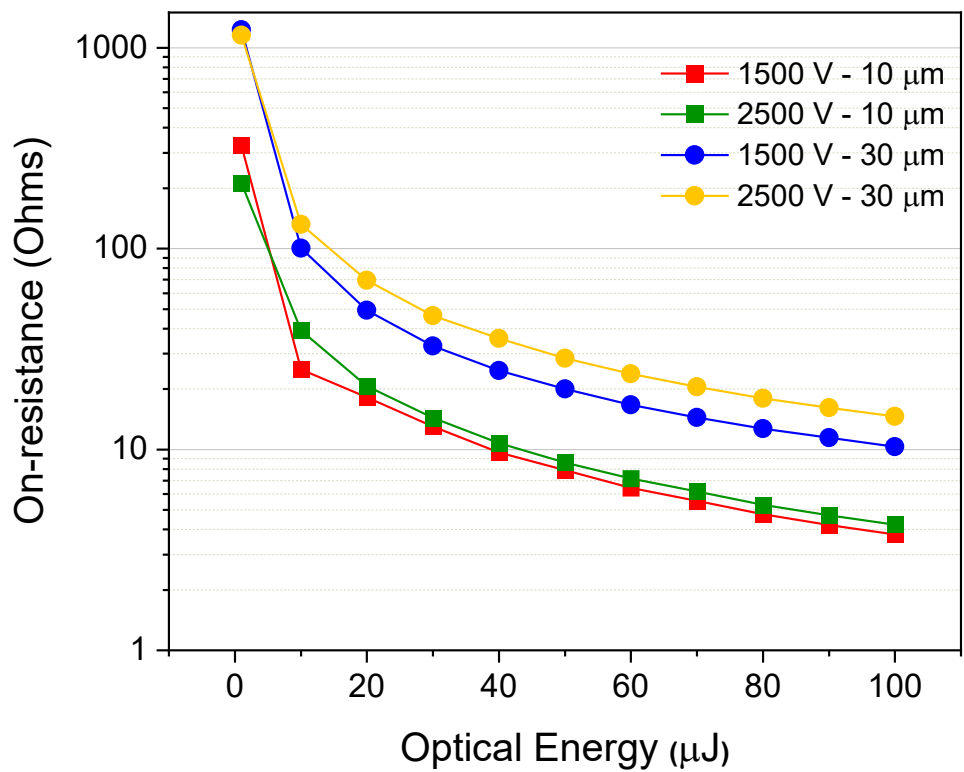

Fig. 4. PCSS on-resistance versus optical energy.

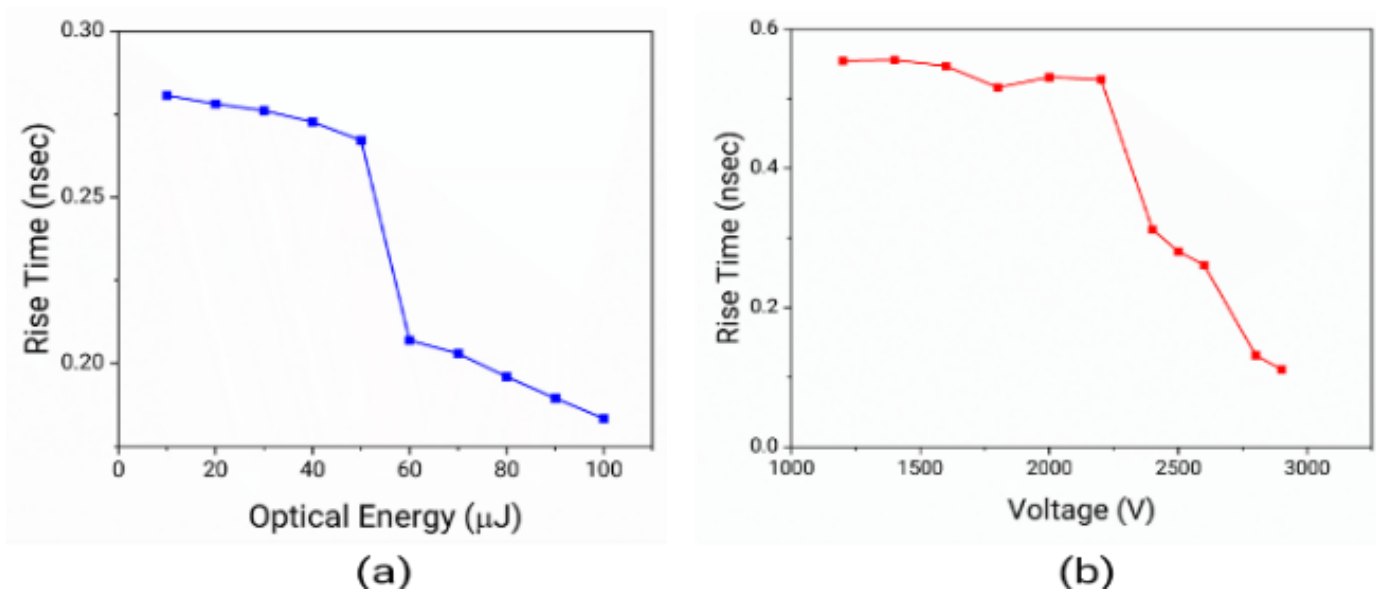

Fig. 5. (a) Rise time with optical energy at 2500 V. (b) Rise time with voltage at 10 μJ optical energy.

So far, uniform illumination has been used across the contact separation. In Fig. 6(a), FG-PCSS photocurrent with a 10 µm contact separation is plotted as a function of beam position (BP) and its pulse width (PW) for 5 different scenarios A-E which are shown in Fig. 6(b). Total optical energy (for a 450 nm wavelength) is kept constant at 10 µJ, and the bias voltage is kept at 1500 V or 2500 V for all cases. Depending on how the beam is positioned, the generated carriers are expected to travel across different distances to reach the anode/cathode contacts and hence, the recombination rate and transit time for these carriers vary. As the optical energy is fixed, at lower PW, a greater number of photons are generated over a smaller area, which, aided with high electric field, onsets the avalanche mode. The BP and PW are both able to produce different photocurrents. The ideal BP for both 1500 V and 2500 V is near the anode with a PW of 2.5 µm (i.e., scenario D). In this case electrons will be collected immediately, while holes gain sufficient energy to reach the cathode with less recombination.

## V. Conclusions

Owing to the high critical electric field of $Ga_2O_3$ compared to GaN and SiC, FG-PCSS yields a smaller contact separation for a given voltage bias. It is demonstrated that with increasingly high voltage bias (yielding higher electric field), FG-PCSS exhibits progressively higher photocurrent (and hence higher device gain) even at low optical energy due to high impact ionization. Increasing the optical energy alone does not change this nonlinear response of the device even though the photocurrent increases. It is also demonstrated that, because of the reduced contact separation, the on-resistance of FG-PCSS is also reduced that requires optical beam with lower energy that enables the possibility of using low-cost and ultra-fast sub-bandgap laser. Finally, it is found that the photocurrent of the FG-PCSS varies with the beam position and width and the applied voltage bias.

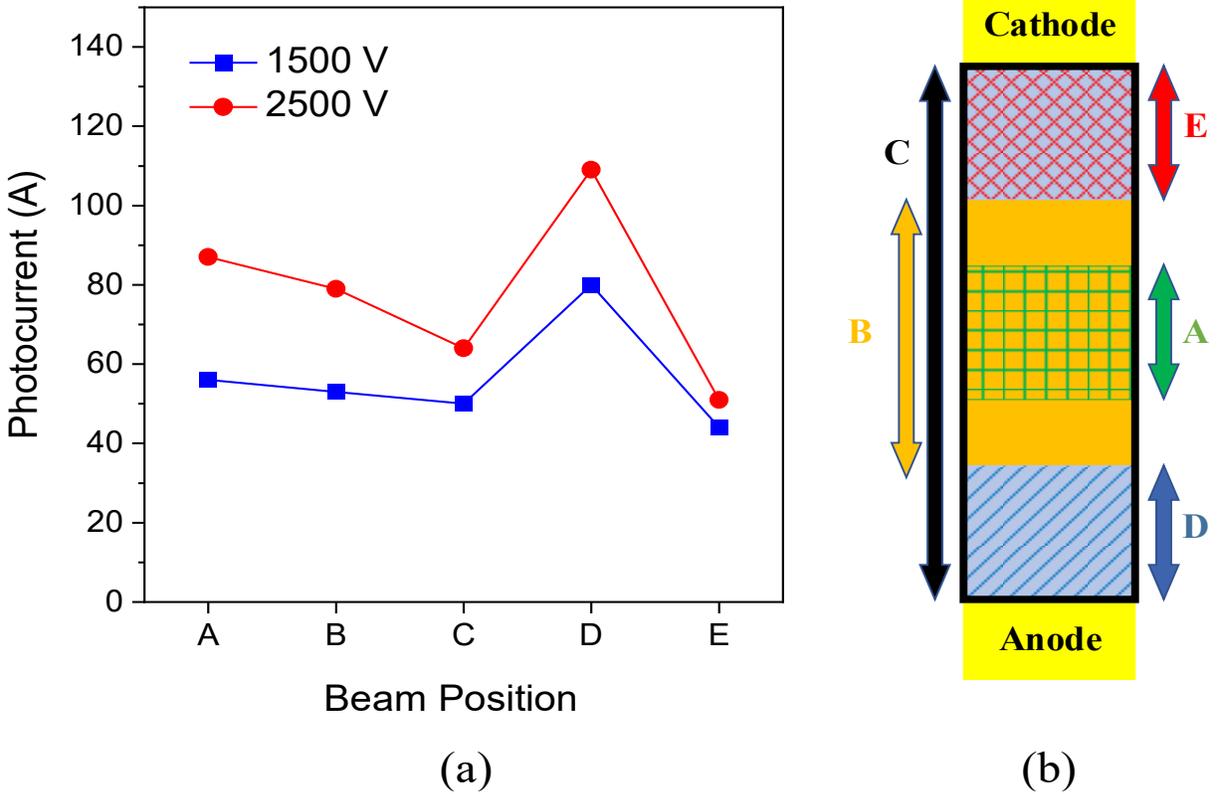

Fig. 6. FG-PCSS (a) photocurrent as a function of beam position (BP) and pulse width (PW) corresponding to 5 scenarios, (b) beam position. A) BP: 5 μm, PW = 2.5 μm, B) BP: 5 μm, PW = 5 μm, C) BP: 5 μm, PW = 10 μm, D) BP: 0 μm PW = 2.5 μm, and E) BP: 10 μm, PW = 2.5 μm.